\documentclass[a4paper,12pt]{article}

\newcommand{\sect}[1]{\setcounter{equation}{0}\section{#1}}

\begin{document}
\topmargin 0pt
\oddsidemargin 0mm

\renewcommand{\thefootnote}{\fnsymbol{footnote}}
\begin{titlepage}
\begin{flushright}
OU-HET 335 \\
hep-th/9912013
\end{flushright}

\vspace{10mm}
\begin{center}
{\Large\bf Surface counterterms and boundary stress-energy tensors
for asymptotically non-anti-de Sitter spaces}
\vspace{15mm}

{\large
Rong-Gen Cai\footnote{e-mail address: cai@het.phys.sci.osaka-u.ac.jp} and
Nobuyoshi Ohta\footnote{e-mail address: ohta@phys.sci.osaka-u.ac.jp}} \\
\vspace{10mm}
{\em Department of Physics, Osaka University,
Toyonaka, Osaka 560-0043, Japan}

\end{center}
\vspace{15mm}
\centerline{{\bf{Abstract}}}
\vspace{5mm}

For spaces which are not asymptotically anti-de Sitter where the
asymptotic behavior is deformed by replacing the cosmological constant
by a dilaton scalar potential, we show that it is possible to have
well-defined boundary stress-energy tensors and finite Euclidean actions
by adding appropriate surface counterterms. We illustrate the method by
the examples of domain-wall black holes in gauged supergravities,
three-dimensional dilaton black holes and topological dilaton black holes
in four dimensions. We calculate the boundary stress-energy tensor and
Euclidean action of these black configurations and discuss their
thermodynamics. We find new features of topological black hole
thermodynamics.

\end{titlepage}

\newpage
\renewcommand{\thefootnote}{\arabic{footnote}}
\setcounter{footnote}{0}
\setcounter{page}{2}


\sect{Introduction}

In the traditional Euclidean path integration approach to black hole
thermodynamics~\cite{Gibbons,Horowitz}, except for the usual
Gibbons-Hawking surface term which makes the variation principle
well defined, one has to choose a suitable reference background and
make subtraction in order to get a finite Euclidean action of black holes.
However, the background subtraction procedure makes the action of black
holes depend on the choice of reference background. Furthermore sometimes
one may encounter the situations in which there are no appropriate reference
backgrounds, as observed for the Taub-NUT-AdS and Taub-Bolt-AdS
spaces~\cite{Cham1,Hawking}. On the other hand, in the quasilocal
formulation of gravity~\cite{Brown1,Brown2}, one can define the so-called
quasilocal stress-energy tensor and conserved charges on the  boundary of
a given spacetime region. Unfortunately, such quantities often diverge as
the boundary is taken to infinity. A suitable background subtraction must
then be made for getting a finite result.

In the asymptotically anti-de Sitter spacetimes, the above difficulty has
been solved recently. The proposal is that by adding suitable surface
counterterms to the gravitational action, one can obtain a well-defined
boundary stress-energy tensor and a finite Euclidean action for the black
hole spacetimes~\cite{Vijay}. A remarkable feature of this procedure is that
the boundary stress-energy tensor and Euclidean action thus defined are
independent of the reference background and the results are physically
unique. Recently a lot of works have been devoted to this proposal and
related topics~\cite{Myers}-\cite{NO}.
In $(n+1)$-dimensional Einstein gravity with a negative cosmological constant
$\Lambda =-n(n-1)/2l^2$, the action can be written as
\begin{equation}
\label{1e1}
S =\frac{1}{16\pi G}\int _{\cal M}d^{n+1}x\sqrt{-g}\left(R
 +\frac{n(n-1)}{l^2}\right)
 -\frac{1}{8\pi G}\int_{\cal \partial M}d^n x \sqrt{-h}K,
\end{equation}
where the first term is called the bulk action, and the second term is
just the Gibbons-Hawking surface term. Here $h$ denotes the reduced metric of
a timelike boundary ${\cal \partial M}$ and $K$ represents the trace of its
extrinsic curvature to be defined below. In ref.~\cite{Emp}, an expression
of surface counterterms has been given, which can cancel divergences up to
$n \le 6$:
\begin{equation}
\label{1e2}
S_{\rm ct}=-\frac{1}{8\pi G}\int_{\cal \partial M}d^nx \sqrt{-h}
 \left[ \frac{n-1}{l} +\frac{l}{2(n-2)}{\cal R}+
 \frac{l^3}{2(n-4)(n-2)^2}\left( {\cal R}_{ab}{\cal R}^{ab}
 - \frac{n}{4(n-1)}{\cal R}^2\right) \right],
\end{equation}
where $\cal R$ and ${\cal R} _{ab}$ are the Ricci scalar and Ricci tensor
for the boundary metric $h_{ab}$. The authors of~\cite{Deb} claimed that
they have given the surface counterterms up to $n\le 8$. From (\ref{1e2}),
one may see that the cosmological constant $l$ plays a crucial role in this
surface counterterm method. Once given the surface counterterms, one may
define a  quasilocal stress-energy tensor by
\begin{equation}
\label{1e3}
T_{ab}=\frac{1}{8\pi G}\left[K_{ab}-Kh_{ab} +\frac{2}{\sqrt{-h}}
 \frac{\delta S_{\rm ct}}{\delta h_{ab}}\right],
\end{equation}
where the extrinsic curvature is $ K_{ab}=-\frac{1}{2}(\Delta_an_b
+\Delta_bn_a)$, and $n_a$ denotes the outward pointing normal vector to
the boundary ${\cal \partial M}$.

Decomposing the boundary metric $h_{ab}$ in the ADM form with a spacelike
surface ${\cal B}$ in ${\cal \partial M}$ with metric $\sigma_{ij}$:
\begin{equation}
\label{1e4}
h_{ab}dx^adx^b=-N_{\cal B}^2dt^2 +\sigma_{ij}(dx^i +N^idt)(dx^j +N^jdt),
\end{equation}
one can define a conserved charge
\begin{equation}
\label{1e5}
Q_{\xi}=\int_{\cal B }d^{n-1}x\sqrt{\sigma}(u^aT_{ab}\xi^b),
\end{equation}
associated with a Killing vector $\xi^a$, where $u^a$ is a timelike unit
normal to ${\cal B}$. In this way one can have the definition of the
mass of gravitational field as~\cite{Vijay}
\begin{equation}
\label{1e6}
M=\int _{\cal B}d^{n-1}x\sqrt{\sigma}N_{\cal B} u^au^bT_{ab}.
\end{equation}
Using the above prescription, Balasubramanian and Kraus~\cite{Vijay} have
obtained the boundary stress tensor associated with a gravitational
system in asymptotically anti-de Sitter space. Via the
AdS/CFT correspondence~\cite{Mald1,Witten,Gubser},
the result is interpreted as the expectation value of the stress tensor of
the boundary quantum conformal field theory. In particular, they have found
a nonvanishing ground state energy for a global $AdS_5$, and have matched
this energy with the Casimir energy of the dual ${\cal N}$=4 super Yang-Mills
theory on $R\times S^3$.

So far, however, most of these works are restricted to the asymptotically
anti-de Sitter space and its asymptotically flat limit. It is true that
these two kinds of spacetimes are much of physical interest, but
there are also other interesting spacetimes which are neither asymptotically
anti-de Sitter nor asymptotically flat. For instance, the geometry in
the decoupling limit of the black D3-brane with NS $B$ field~\cite{Mald2,HI}
has been proposed as the gravity dual of the ${\cal N}$=4 super
Yang-Mills theory in non-commutative spacetime. It is thus interesting to
try to extend this approach to a more general class of spacetimes. In this
paper we consider the kind of spacetimes which are not asymptotically
anti-de Sitter, in which the asymptotically anti-de Sitter behavior is
deformed by the presence of a dilaton potential in the bulk action. In this
class of spacetimes, we find that it is also possible to have a well-defined
boundary stress-energy tensor and a finite Euclidean action by slightly
modifying the above prescription. We give a general form of the surface
counterterms necessary to cancel the divergences and provide a formula for
the coefficient in terms of the asymptotic behaviors of the metrics and
potential in the solution.

The organization of this paper is as follows. In the next section we first
consider domain-wall spacetimes in gauged supergravities which come from
the sphere reduction of D$p$-branes in type II supergravities, since
in this case  we can have a consistency check of our result. We will also
consider a kind of charged domain-wall spacetimes. Our results can be
regarded as a part of the realization of the so-called domain
wall/QFT correspondence.  In Sec.~3 we will discuss a three-dimensional
dilaton black hole, where the BTZ black hole is deformed by a dilaton
potential. In Sec.~4 we extend this discussion to the topological dilaton
black holes in four dimensions. A brief summary is given in Sec.~5.
The general formula for the surface counterterms and their coefficients
are summarized in the appendix.


\sect{ Domain-wall black holes}

The AdS/CFT correspondence asserts that there is an equivalence between
the bulk supergravity (string/M theories) and a boundary conformal field
theory. This correspondence nicely illustrates the holographic principle
\cite{tHooft,Susskind} which is widely believed to be a feature of any
consistent theory of quantum gravity. Thus the AdS/CFT correspondence
is just a special case of a more general correspondence between supergravity
and quantum field theory (QFT) in one lower dimensions. On the basis of
the observation that the AdS metric in horoshperical coordinates is a
special case of a domain-wall metric, Boonstra, Skenderis, and
Townsend~\cite{BST} have extended the AdS/CFT correspondence to the
so-called domain-wall/QFT correspondence between the gauged supergravity
and quantum field theory on domain walls. This correspondence has been
discussed further in ref.~\cite{Beh} in various dimensions. It is
straightforward to extend this to the correspondence between the domain-wall
black holes and corresponding quantum field theory at finite temperature.
In this section we will extract the stress-energy tensor of quantum field
theory on the domain-walls in the spirit of domain-wall/QFT
correspondence. We first discuss the case, in which the domain-wall black
holes come from sphere reduction of D$p$-branes in the ``dual'' frame.
We will then consider the charged domain-wall black holes
which come from singular sphere reductions of eleven-dimensional supergravity
and ten-dimensional type IIB supergravity.

\subsection{Neutral domain-wall black holes}

Let us consider the black D$p$-brane solution with ``magnetic'' charge in
type II supergravity. In the string frame, the action is
\begin{equation}
\label{2e1}
S=\frac{1}{16\pi G_{10}}\int d^{10}x\sqrt{-g} \left[ e^{-2\phi}
 \left(R +4(\partial \phi)^2\right) -\frac{1}{2(8-p)!}F^2_{8-p}\right],
\end{equation}
where $G_{10}=8\pi^6 \alpha'^4$ is the gravitational constant in
ten dimensions. The black D$p$-brane solution is
\begin{eqnarray}
\label{2e2}
ds^2_{\rm string} &=& H^{-1/2}(-fdt^2 +dx_p^2) + H^{1/2}(f^{-1}dr^2 +
 r^2 d\Omega_{8-p}^2), \nonumber \\
e^{\phi} &=& g_s H^{(3-p)/4}, \nonumber \\
F_{8-p} &=& Q\epsilon_{8-p},
\end{eqnarray}
where $g_s$ is the string coupling constant, $\epsilon_{8-p}$ the volume
form of $S^{8-p}$ and
\begin{equation}
H=1 +\frac{r^{7-p}_0\sinh^2\alpha}{r^{7-p}}, \ \
f=1-\left(\frac{r_0}{r}\right)^{7-p}.
\end{equation}
In the decoupling limit: $\alpha' \rightarrow 0$, but keeping fixed
$ U=r/\alpha', U_0=r_0/\alpha'$ and the 't Hooft coupling constant
$g^2_{\rm YM}N$, with $g^2_{\rm YM} = g_s (\alpha')^{(p-3)/2}$, the
harmonic function tends to
\begin{equation}
H=1+\frac{g^2_{\rm YM}N}{(\alpha')^2U^{7-p}} \; \Longrightarrow \;
 g^2_{\rm YM}N(\alpha')^{-2}U^{p-7},
\end{equation}
where we have absorbed an unimportant coefficient into
$g_{\rm YM}$~\cite{BST}. Except for the case of $p=3$, the radius of angle
part of the string metric (\ref{2e2}) depends on  $U$ and the metric is
singular at $U=0$ even in the case of $U_0=0$. To circumvent this problem,
the so-called ``dual frame'' metric has been considered in \cite{BST}:
\begin{equation}
ds^2_{\rm dual} =(Ne^{\phi})^{2/(p-7)}ds^2_{\rm string}.
\end{equation}
In this frame, the action (\ref{2e1}) becomes
\begin{equation}
S=\frac{N^2}{16\pi G_{10}}\int d^{10}x\sqrt{-g}(Ne^{\phi})^{\lambda}
 \left [R +\frac{4(p-1)(p-4)}{(7-p)^2}(\partial \phi)^2
 - \frac{1}{2N^2(8-p)!}F^2_{8-p}\right],
\end{equation}
where
\begin{equation}
\lambda =2(p-3)/(7-p).
\label{c1}
\end{equation}
The decoupling limit solution in the ``dual frame'' is
\begin{eqnarray}
\label{2e7}
ds^2_{\rm dual} &=& \alpha' \left[\left(g^2_{\rm YM}N\right)^{-1}U^{5-p}
 (-fdt^2+dx_p^2) +U^{-2}f^{-1}dU^2 +d\Omega^2_{8-p}\right], \nonumber \\
e^{\phi} &=&\frac{1}{N}\left[\left(g^2_{\rm YM}N\right)
 U^{p-3}\right]^{(7-p)/4}, \nonumber\\
F_{8-p} &=& (7-p)N (\alpha')^{(7-p)/2}\epsilon_{8-p},
\end{eqnarray}
where $f=1- (U_0/U)^{7-p}$. The near-horizon ``dual frame'' metric is
$AdS_{p+2}\times S^{8-p}$ for $p\ne 5$ and $E^{(1,6)}\times S^3$ for $p=5$.
An important feature of this frame is that the radius of the angle part
of the metric becomes a constant.

Because $\alpha'$ is eventually canceled at the end of the calculations,
we set $\alpha'=1$ in what follows. In addition,
by the transformation ($p\ne 5$)
\begin{equation}
u^2 =\Re ^2 \left(g^2_{\rm YM}N\right)^{-1}U^{5-p}, \ \ \Re=2/(5-p),
\label{c2}
\end{equation}
the above ``dual frame'' metric can be put in a standard form
\begin{eqnarray}
ds^2_{\rm dual} &=& \frac{u^2}{\Re^2}\left (-fdt^2 +dx_p^2 \right)
 +\frac{\Re^2}{u^2 f}du^2 +d\Omega^2_{8-p}, \nonumber\\
e^{\phi}&=& \frac{1}{N}\left(g^2_{\rm YM}N\right)^{(7-p)/2(5-p)}
 (u/\Re)^{(p-7)(p-3)/2(p-5)}, \nonumber \\
F_{8-p} &=& (7-p)N \epsilon_{8-p}.
\label{c3}
\end{eqnarray}
It was found that the scale $u$ introduced above is just the holographic
energy scale of the boundary QFT. Thus the ``dual frame'' was argued as the
holographic frame describing supergravity probes \cite{BST}.

Due to the fact that the radius of angle part of the metric is a constant,
one may consistently reduce the angle part to get an effective gauged
$(p+2)$-dimensional supergravity. In the Einstein frame, the
resulting action is
\begin{equation}
\label{2e10}
S=\frac{N^2 \Omega_{8-p}}{(2\pi)^7 }\int d^{p+2}x\sqrt{-g}
 \left[R-\frac{1}{2}(\partial \Phi)^2 +V(\Phi)\right]-\frac{2N^2\Omega_{8-p}}
 {(2\pi)^7 }\int d^{p+1}x\sqrt{-h}K,
\end{equation}
where for the later use, we have added the Gibbons-Hawking term to the
bulk action, $\Omega_{8-p}$ is the volume of a unit $(8-p)$-sphere, and
\begin{eqnarray}
\label{2e11}
&& V(\Phi)=\frac{1}{2}(9-p)(7-p)N^{-2\lambda/p}e^{a\Phi}, \nonumber\\
&&\Phi = \frac{2\sqrt{2(9-p)}}{\sqrt{p}(7-p)}\phi, \ \
a = -\frac{\sqrt{2}(p-3)}{\sqrt{p(9-p)}}.
\end{eqnarray}
In the effective gauged supergravity action, its equations of motion
have the following domain-wall black hole solutions:
\begin{eqnarray}
\label{2e12}
ds^2_{p+2} &=& (Ne^{\phi})^{2\lambda/p}\left[\frac{u^2}{\Re ^2}
 \left(-\tilde{f} dt^2 +dx_p^2 \right) +\frac{\Re^2}
 {u^2 \tilde{f}}du^2 \right], \nonumber\\
\tilde{f} &=& 1-\left(\frac{u_0}{u}\right)^{2(7-p)/(5-p)},
\end{eqnarray}
where $\lambda, \Re$ and $e^{\phi}$ are given in Eqs.~(\ref{c1}), (\ref{c2})
and (\ref{c3}), respectively, and $u_0$ is defined as $u_0^2 =\Re^2
(g^2_{\rm YM}N)^{-1}U_0^{5-p}$.

Now we are interested in extracting the stress-energy tensor of quantum field
which lives in the domain wall (\ref{2e12}), according to the
domain-wall/QFT correspondence. We find that the scalar potential occurring
in the action (\ref{2e10}) can play the same role as a cosmological constant
does in the asymptotically anti-de Sitter spaces. Writing the scalar
potential $V(\Phi)$ as
\begin{equation}
V(\Phi)\equiv \frac{n (n-1)}{l^2_{\rm eff}} =\frac{p(p+1)}{l^2_{\rm eff}},
\end{equation}
one may introduce an ``effective cosmological constant'' $1/l_{\rm eff}$
defined by
\begin{equation}
\label{2e15}
\frac{1}{l_{\rm eff}}=\sqrt{\frac{V(\Phi)}{p(p+1)}}.
\end{equation}
According to the formulae (\ref{for1}), (\ref{for2}) and (\ref{a1}) in the
appendix, by adding the following surface counterterm to (\ref{2e10}):
\begin{equation}
S_{\rm ct}=-\frac{2N^2 \Omega_{8-p}}{(2\pi)^7}
  \int d^{p+1}x \sqrt{-h}\frac{c_0}{l_{\rm eff}}, \ \
c_0=\sqrt{\frac{(9-p)p(p+1)}{2(7-p)}},
\label{b1}
\end{equation}
it is possible to cancel divergences in physical quantities such as
stress-energy tensor and Euclidean action associated with the domain-wall
black holes. Note that the surface term is similar to the first term in
(\ref{1e2}) but with different coefficient.

Using (\ref{1e3}) we have
\begin{equation}
\label{2e17}
T_{ab}=\frac{2N^2 \Omega_{8-p}}{(2\pi)^7}\left [K_{ab}-Kh_{ab}
 -\frac{c_0}{l_{\rm eff}}h_{ab}\right],
\end{equation}
where the labels $a, b$ run over the domain-wall directions. Substituting
the solution (\ref{2e12}) into (\ref{2e17}) and using the `` effective
cosmological constant'' (\ref{2e15}), we obtain
\begin{eqnarray}
\label{2e18}
&& \frac{(2\pi)^7}{2N^2\Omega_{8-p}}T_{tt}=\frac{9-p}{4}
 \left(g^2_{\rm YM}N\right)^{(p-3)/p(5-p)}\left(\frac{u}{\Re}\right)
 ^{(p^2-4p-9)/p(p-5)}\left(\frac{u_0}{u}\right)^{2(7-p)/(5-p)}
 + \cdots, \nonumber \\
&& \frac{(2\pi)^7}{2N^2\Omega_{8-p}}T_{ij}=\delta_{ij}\frac{5-p}{4}
 \left(g^2_{\rm YM}N\right)^{(p-3)/p(5-p)}\left (\frac{u}{\Re}\right)
 ^{(p^2-4p-9)/p(p-5)}\left(\frac{u_0}{u}\right)^{2(7-p)/(5-p)}
 + \cdots, \nonumber \\
\end{eqnarray}
where dots denote higher order terms, which will vanish when we take
the boundary to the spatial infinity. Using (\ref{1e6}) we get the mass
of the domain-wall black hole,
\begin{eqnarray}
\label{2e19}
M &=&\int_{u\to \infty}d^px \sqrt{\sigma}N_{\cal B} u^tu^t T_{tt} \nonumber\\
  &=& \frac{\Omega_{8-p}}{(2\pi)^7 g^4_{\rm YM}}\frac{9-p}{2}U_0^{7-p}V_p,
\end{eqnarray}
where $V_p$ is the spatial volume of the domain wall.

The surface metric $\gamma_{ab}$ of the spacetime, in which the
boundary quantum field lives, can be obtained as \cite{Myers}
\begin{equation}
\label{2e20}
\gamma_{ab} =\lim_{u \rightarrow \infty} \frac{\Re^2}{u^2}
 (Ne^{\phi})^{-2\lambda/p}h_{ab}
=\eta_{ab},
\end{equation}
which means that the boundary quantum field theory lives in a flat domain
wall. The stress-energy tensor of boundary quantum field theory can be
obtained as \cite{Myers}
\begin{equation}
\label{2e21}
\sqrt{-\gamma}\gamma^{ab}\tau_{bc}= \lim_{u\to \infty}
 \sqrt{-h}h^{ab}T_{bc}.
\end{equation}
Substituting (\ref{2e18}) into the above formula, we finally arrive at
\begin{equation}
\label{2e22}
\tau_{ab} =\frac{U_0^{7-p}\Omega_{8-p}}{(2\pi)^7 g^4_{\rm YM}}\;
{\rm diag} \left[\frac{9-p}{2}, \frac{5-p}{2},\cdots,\frac{5-p}{2}\right],
\end{equation}
which can be interpreted as the vacuum expectation value of the quantum field
theory on the domain wall (\ref{2e12}).

On the other hand, we can calculate the Euclidean action of the
domain-wall black hole
\begin{eqnarray}
\label{2e23}
I= &-& \frac{N^2 \Omega_{8-p}}{(2\pi)^7 }\int d^{p+2}x\sqrt{g}
 \left[R-\frac{1}{2}(\partial \Phi)^2 +V(\Phi)\right] \nonumber \\
&+& \frac{2N^2\Omega_{8-p}} {(2\pi)^7 }\int d^{p+1}x\sqrt{h}K
 + \frac{2N^2 \Omega_{8-p}}{(2\pi)^7}
 \int d^{p+1}x \sqrt{h}\frac{c_0}{l_{\rm eff}},
\end{eqnarray}
where the first line is the bulk contribution, the second term is the usual
Gibbons-Hawking surface term, and the last is just the surface counterterm.
Putting the solution into (\ref{2e23}) yields a finite result
\begin{equation}
\label{2e24}
I=-\frac{\Omega_{8-p}V_pU^{7-p}_0}{(2\pi)^7 g^4_{\rm YM}T}
 \frac{5-p}{2},
\end{equation}
from which we obtain the free energy ${\cal F}$
\begin{equation}
\label{2e25}
{\cal F}\equiv TI =
-\frac{ \Omega_{8-p}}{(2\pi)^7 g^4_{\rm YM}}
 \frac{5-p}{2}V_pU_0^{7-p},
\end{equation}
where $T$ is the Hawking temperature of the domain-wall black hole
(\ref{2e12}). For $p<5$, the free energy is negative, which implies that
the system is thermodynamically stable, while for $p>5$, the free energy
becomes positive. In this case, the thermal excitations are thermodynamically
unstable, and they will be suppressed in canonical ensemble. From
(\ref{2e12}) it might appear that the results derived above are applicable
only to $p<5$. In fact, the above results hold for $p\ge 5$ as well, for
we may use the coordinate $U$ in (\ref{2e7}) instead of $u$ in (\ref{2e12})
and then the same results are obtained (the apparent singularities for $p=5$
as in Eqs.~(\ref{2e18}) are absent in terms of $U$ and $U_0$). We also see
from (\ref{2e22}) and (\ref{2e25}) that the case of $p=5$ is a bit peculiar:
the free energy and the pressure of thermal excitations on the domain wall
vanish.

\subsection{A consistency check}

Note that the domain-wall black hole solution (\ref{2e12}) comes from
the sphere reduction of D$p$-brane solution (\ref{2e2}). It makes
possible to calculate the stress-energy tensor of boundary quantum field
theory and the free energy directly from the D$p$-brane solution
(\ref{2e2}) and to compare with the results from the counterterm method.
The formula to extract the stress-energy tensor of excitations of D$p$-branes
has been given in \cite{Myers}:
\begin{equation}
\label{2e26}
T_{ab}=\frac{1}{16\pi G_{10}g_s^2}\int _{r\to \infty}
    d\Omega_{8-p}r^{8-p} n^i[\eta_{ab}(\partial_ih^c_{\ c}
    +\partial_i h^j_{\ j} -\partial_jh^j_{\ i})-\partial_ih_{ab}],
\end{equation}
where $n^i$ is a radial unit in the transverse subspace, while
$h_{\mu\nu}=g_{\mu\nu}-\eta_{\mu\nu}$ is the deviation of the (Einstein
frame) metric from  that  for flat space. The labels $a, b=0, 1,\cdots, p$
run over the world-volume directions, while $i, j =1,2, \cdots, 9-p$ denote
the transverse directions. In addition, it should be reminded that the
calculations in (\ref{2e26}) must be done using asymptotically Cartesian
coordinates.

Rewriting the D$p$-brane solution (\ref{2e2}) in the isotropic coordinates
of the  Einstein frame, we have
\begin{equation}
\label{2e27}
ds^2_{\rm E}=H^{-(7-p)/8}(-fdt^2 +dx_p^2) +H^{(p+1)/8}r^2 \rho^{-2}
   (d\rho^2 +\rho^2d\Omega^2_{8-p}),
\end{equation}
where $\rho$ is the radial coordinate having the relation with $r$ as
\begin{equation}
r^{7-p}=\rho^{7-p}\left (1+\frac{r_0^{7-p}}{4\rho^{7-p}}\right)^2.
\end{equation}
Substituting the solution into (\ref{2e26}), one finds
\begin{equation}
\label{2e29}
T_{ab}=\frac{(7-p)r_0^{7-p}\Omega_{8-p}}{16\pi G_{10}g_s^2}\;
   {\rm diag} \left [\frac{8-p}{7-p}+\sinh^2\alpha,
      \frac{1}{7-p}-\sinh^2\alpha,\cdots,
      \frac{1}{7-p}-\sinh^2\alpha \right].
\end{equation}
This stress-energy tensor includes the contribution from the extremal
background, which can be obtained from (\ref{2e29}) by taking $r_0
\rightarrow 0$, but keeping $\tilde{R}^{7-p}= r_0^{7-p}
\sinh\alpha\cosh\alpha $ constant:
\begin{equation}
(T_{ab})_{\rm ext}=
\frac{(7-p)\Omega_{8-p}}{16\pi G_{10}g_s^2}\;
   {\rm diag} \; [\tilde{R}^{7-p},
     -\tilde{R}^{7-p},\cdots,
      -\tilde{R}^{7-p}].
\end{equation}
Subtracting the contribution of extremal background from (\ref{2e29})
and taking the near-extremal limit: $r^{7-p}_0\sinh^2\alpha
 \approx \tilde{R}^{7-p} -r_0^{7-p}/2$, we reach
\begin{equation}
\label{2e31}
(\triangle T)_{ab}=\frac{\Omega _{8-p}r_0^{7-p}}{16\pi G_{10}g_s^2}
   \frac{1}{2}\; {\rm diag}\; [9-p,5-p,\cdots,5-p],
\end{equation}
and its trace
\begin{equation}
\triangle T= -\frac{(p-3)^2}{2}\frac{\Omega_{8-p}r_0^{7-p}}
      {16\pi G_{10}g_s^2}.
\end{equation}
In the decoupling limit, (\ref{2e31}) reduces to
\begin{equation}
\label{2e33}
(\triangle T)_{ab}=\frac{\Omega _{8-p}U_0^{7-p}}{(2\pi)^7 g^4_{\rm YM}}
   \frac{1}{2}\; {\rm diag}\; [9-p,5-p,\cdots,5-p],
\end{equation}
which precisely agrees with (\ref{2e22}) obtained by the counterterm
method. In addition, from the 00-component of $(\triangle T)_{ab}$ we can
read off directly the energy of thermal excitations on the D$p$-branes:
\begin{equation}
E= \frac{\Omega_{8-p}}{(2\pi)^7 g^4_{\rm YM}}\frac{9-p}{2}U_0^{7-p}V_p.
\end{equation}
Obviously it is again the same as the mass (\ref{2e19}) of the domain-wall
black holes.

For the black D$p$-brane (\ref{2e2}), the Hawking temperature and
entropy are
\begin{eqnarray}
&& T=\frac{1}{4\pi}\frac{7-p}{r_0\cosh\alpha}, \nonumber\\
&& S=\frac{4\pi \Omega_{8-p}V_p}{(2\pi)^7g_s^2}r_0^{8-p}\cosh\alpha.
\end{eqnarray}
The free energy of the thermal excitations defined as ${\cal F}=E-TS$ is
\begin{equation}
{\cal F}= -\frac{ \Omega_{8-p}}{(2\pi)^7 g^4_{\rm YM}}
   \frac{5-p}{2}V_p U^{7-p}_0,
\end{equation}
in the decoupling limit. Once again, this reproduces the result
(\ref{2e25}) by the surface counterterm method.

\subsection{Charged domain-wall black holes}

It is now clear that one can make consistent reductions of
eleven-dimensional supergravity on $S^4$ or $S^7$, and ten-dimensional
type IIB supergravity on $S^5$. The Kaluza-Klein sphere reduction results
in gauged supergravities. The anti-de Sitter spaces are vacuum solutions
of these gauged supergravity. More recently an evidence has been
provided that some singular limits of sphere reduction are also
consistent and resulting gauged supergravities have domain-wall vacuum
solutions, instead of the AdS spaces \cite{Cvetic}.

The so-called domain-wall supergravities can be consistently truncated
to the following bosonic Lagrangian \cite{Cvetic}:
\begin{equation}
\label{2e37}
S=\frac{1}{16\pi G_{p+2}}\int d^{p+2}x\sqrt{-g}\left [
      R-\frac{1}{2}(\partial \phi)^2-\frac{1}{4}e^{a\phi}F_{\mu\nu}
      F^{\mu\nu} +\frac{1}{2} b^2 e^{-a\phi}\right],
\end{equation}
where $a^2=2/p$,  $ b$ is a constant and $F_{\mu\nu}$ denotes
the Maxwell field strength. The action (\ref{2e37}) for $p=5$
comes from the reduction on $S^3\times R$ of eleven-dimensional
supergravity, while $p=2$ from the reduction on $S^3\times R^4$ and $p=3$
from the $ S^3\times R^2$ reduction of type IIB supergravity. But we
consider an arbitrary $p$ in what follows.

The equations of motion from  the action (\ref{2e37}) have domain-wall
black hole solutions
\begin{eqnarray}
&& ds^2 =-f(r) dt^2 +f^{-1}(r) dr^2 +r dx_p^2 ,\nonumber\\
&& F_{tr}=\frac{pq}{2}r^{-(p+2)/2}, \ \ \  e^{a\phi}=r, \nonumber\\
&& f(r)= 2r\left (\frac{b^2}{p^2}+\frac{q^2}{4r^p}-\frac{m}{r^{p/2}}
     \right),
\end{eqnarray}
where $m$ and $q$ represent two integration constants.
Choosing the surface counterterm as
\begin{equation}
S_{\rm ct}=-\frac{1}{8\pi G_{p+2}}\int d^{p+1}x\sqrt{-h} \frac{c_0}
  {l_{\rm eff}},\ \ \ c_0=\sqrt{p(p+1)},\ \ \
  \frac{1}{l_{\rm eff}}=b e^{-a\phi/2}\sqrt{\frac{1}{2p(p+1)}},
\label{b2}
\end{equation}
as prescribed in the appendix, we have the quasilocal stress-energy tensor
\begin{eqnarray}
&& 8\pi G_{p+2} T_{tt}=\frac{bm}{\sqrt{2}}r^{(1-p)/2} +\cdots, \nonumber\\
&& 8\pi G_{p+2} T_{ij}=0 +{\cal O}(r^{-(p+1)/2}) +\cdots.
\end{eqnarray}
The mass of the black hole can be obtained as follows:
\begin{equation}
M=\int_{r\rightarrow \infty}d^pxr^{p/2}f^{-1/2}T_{tt}
   =\frac{pmV_p}{16\pi G_{p+2}}.
\end{equation}
In this case, the surface metric $\gamma_{ab}$ is
\begin{equation}
\gamma_{ab}dx^a dx^b =\lim_{r\to \infty}\frac{1}{r}h_{ab}dx^a dx^b
      =-\frac{2b^2}{p^2}dt^2 +dx_p^2,
\end{equation}
and the boundary stress-energy tensor $\tau_{ab}$ is found to be
\begin{equation}
\tau_{ab}=\frac{bm}{8\pi G_{p+2}\sqrt{2}}\left[1,0,\cdots,0\right].
\end{equation}
Its pressure vanishes identically. This is reminiscent of the case of the
$p=5$ neutral domain-wall black holes in the previous subsection.
Calculating the Euclidean action of the charged domain-wall black holes,
\begin{eqnarray}
\label{2e44}
I= &-&\frac{1}{16\pi G_{p+2}}\int d^{p+2}x\sqrt{g}\left [
      R-\frac{1}{2}(\partial \phi)^2-\frac{1}{4}e^{a\phi}F_{\mu\nu}
      F^{\mu\nu} +\frac{1}{2} b^2 e^{-a\phi}\right] \nonumber \\
   &+&\frac{1}{8\pi G_{p+2}}\int d^{p+1}x\sqrt{h}K
     +\frac{1}{8\pi G_{p+2}}\int d^{p+1}x\sqrt{h}\frac{c_0}
   {l_{\rm eff}},
\end{eqnarray}
we find that the Euclidean action $I$ vanishes identically. This is again
the same as the case of $p=5$ neutral domain-wall black holes. Here it should
be reminded that the calculation (\ref{2e44}) has been done implicitly in
grand canonical ensemble, in which the electric potential of the charge of
black holes is fixed at the boundary.  The vanishing of the
Euclidean action in grand canonical ensemble means that the Gibbs free energy
${\cal G}=0$  for the charged domain-wall black holes.


\sect{Three-dimensional  dilaton black holes}

The three-dimensional BTZ black hole plays an important role in understanding
statistical entropy of black holes. The degrees of freedom of the BTZ black
hole can be accounted for by a two-dimensional boundary conformal field
theory, which is a special case of the AdS/CFT correspondence.
In this section we consider a deformed BTZ black hole by a dilaton field
and an exponential potential. Its action is
\begin{equation}
\label{3e1}
S=\frac{1}{16\pi G} \int d^3x\sqrt{-g}(R-4(\partial\phi)^2 +2\Lambda
       e^{b\phi})-\frac{1}{8\pi G}\int d^2x \sqrt{-h}K,
\end{equation}
where $b$ and $\Lambda$ are two constants. The action has the following
 black hole solution \cite{Chan1,Chan2}:
\begin{eqnarray}
\label{3e2}
&& ds^2 =-A(r)dt^2 +A(r)^{-1}\beta^2 dr^2 +\beta^2 r^Nd\theta^2,
             \nonumber\\
&& \phi = k\ln \ r, \nonumber\\
&& A(r)= -\frac{16 G m}{N}r^{1-N/2} +\frac{8\Lambda \beta^2 }
        {N(3N-2)}r^N, \nonumber\\
&& k=\pm \frac{1}{4}\sqrt{N(2-N)},\ \ \ bk=N-2,
\end{eqnarray}
where $m$ is the quasilocal mass identified at spatial infinity by using
background subtraction. The positive
mass ($m >0$) black holes exist only for $2\ge N >2/3$ and $\Lambda>0$. When
$N=2$, the solution reduces to the BTZ black hole. In addition, note that
the radial coordinate $r$ is chosen to be dimensionless and $\beta$ is a
length scale with dimension of length. (This should not be confused with the
inverse Hawking temperature $1/T$. In this paper we do not use $\beta$ for
the inverse Hawking temperature.)

The dilaton black hole solution (\ref{3e2}) has the horizon at $r=r_+$ with
\begin{equation}
r_+^{(3N-2)/2}=\frac{2Gm(3N-2)}{\Lambda \beta^2}.
\end{equation}
The Hawking temperature and entropy of the solution are
\begin{eqnarray}
&& T =\frac{2Gm(3N-2)}{N\pi \beta}r_+^{-N/2}, \nonumber\\
&& S = \frac{\pi \beta} {2G} r_+^{N/2}.
\end{eqnarray}
Hence we have the free energy of the solution
\begin{equation}
\label{3e5}
{\cal F}=m-TS =-2m(N-1)/N.
\end{equation}
For $2>N>1$, the free energy is always negative and the
dilaton black hole is thermodynamically stable as the BTZ black hole.
For $ 1>N>2/3$, however, the free energy becomes positive and
the dilaton black hole is thermodynamically unstable.

Obviously the dilaton black hole solution (\ref{3e2}) is not asymptotically
anti-de Sitter, unless $N=2$. In  what follows we will extract the boundary
stress-energy tensor and its quantum expectation value of the corresponding
boundary quantum field, by adding an appropriate surface counterterm to
the action ({\ref{3e1}). As in the previous section, the occurrence of the
dilaton potential makes possible to choose a suitable surface counterterm as
\begin{equation}
\label{3e6}
S_{\rm ct} =-\frac{1}{8\pi G}\int d^2x\sqrt{-h}\frac{c_0}{l_{\rm eff}},
 \ \ \ c_0=\sqrt{\frac{2N}{3N-2}},\ \ \ \frac{1}{l_{\rm eff}}=
   \sqrt{\Lambda}e^{b\phi/2},
\label{b3}
\end{equation}
as given in the appendix. Using the quasilocal stress-energy tensor formula
(\ref{1e3}), in this case, we have
\begin{eqnarray}
8\pi G T_{tt} &=& \frac{4G m}{\beta}c
     +\cdots, \nonumber \\
8\pi G T_{\theta\theta} &=& 8 G m\beta \frac{N-1}{Nc}
       +\cdots,
\end{eqnarray}
where $c^2=8\Lambda \beta^2/N(3N-2)$. The mass of the black hole is found to be
\begin{equation}
M=\int_{r\rightarrow \infty}d\theta \beta r^{N/2}A^{1/2}(r)
         u^tu^tT_{tt}=m,
\end{equation}
which means that the mass of the black hole from the counterterm method
is the same as the quasilocal mass identified at the spatial infinity.
Note that the latter is obtained by using background subtraction method
\cite{Brown1,Brown2}.

The surface metric is derived as
\begin{equation}
\gamma_{ab}dx^a dx^b=\lim_{r\to \infty}\frac{1}{r^N}h_{ab}dx^a dx^b
 =-c^2 dt^2+\beta^2 d\theta^2.
\end{equation}
In this spacetime, the boundary stress-energy tensor of the corresponding
quantum field can be calculated as in the previous examples and we get
\begin{equation}
\tau_{ab}=\frac{mc}{2\pi \beta}\left[1,\frac{2(N-1)\beta^2}{Nc^2}\right].
\end{equation}
Furthermore, calculating the Euclidean action of the black hole,
\begin{equation}
I=-\frac{1}{16\pi G} \int d^3x\sqrt{g}(R-4(\partial\phi)^2 +2\Lambda
       e^{b\phi})+\frac{1}{8\pi G}\int d^2x \sqrt{h}K +\frac{1}{8\pi G}
     \int d^2x\sqrt{h}\frac{c_0}{l_{\rm eff}},
\end{equation}
yields a finite result
\begin{equation}
I=\frac{{\cal F}}{T}=-2m\frac{N-1}{NT},
\end{equation}
which gives us the same free energy as (\ref{3e5}). The example of the
three-dimensional dilaton black hole shows that the surface counterterm
method works well as in the case of domain-wall black holes.  We expect
that this method is also applicable to other three-dimensional black holes
with a nonvanishing scalar field.


\sect{Topological dilaton black holes}

Recently it has been found that in the asymptotically anti-de Sitter spaces,
except for the black holes whose horizon hypersurface has the topology
of positive curvature sphere, there are other black hole solutions with
horizon hypersurfaces of zero or negative constant curvature. The latter
are called topological black holes. These topological black holes have been
studied extensively in the AdS/CFT correspondence (for example, see \cite{Em}
and references therein). In this section, we consider those topological
black holes in dilaton gravities. That is, as in the case of
three-dimensional dilaton black holes, the negative cosmological
constant is replaced by a dilaton potential, which changes drastically
the asymptotic behavior of black hole solutions.

The action we will consider is
\begin{equation}
\label{4e1}
S=\frac{1}{16\pi G}\int d^4x\sqrt{-g}\left[ R-2(\partial\phi)^2
    +2\Lambda e^{2b\phi} -e^{-2a\phi}F_{\mu\nu}F^{\mu\nu}\right]
 -\frac{1}{8\pi G}\int d^3x\sqrt{-h}K,
\end{equation}
where $a$ and $b$ are two constants.  Assuming the solution has the
following metric:
\begin{equation}
\label{4e2}
ds^2 =-A(r)dt^2 +A^{-1}(r) dr^2 +R^2(r)d\Sigma_k^2,
\end{equation}
where $d\Sigma^2_k$ is the line element of a two-dimensional
hypersurface with constant curvature $2k$. Without loss of
the generality, we may set $k=1$, $0$ and $ -1$, respectively.
When $k=1$, the hypersurface $\Sigma$ has a positive constant curvature.
This is the case of spherically symmetric black holes. The horizon
surface is of the topology of two-sphere $S^2$. When $k=0$, the hypersurface
$\Sigma$ is a Ricci flat surface. In this case, we may have the two-torus
topology $T^2$, or its infinite area limit $R^2$, or a cylinder
topology $S^1 \times R$. Finally when $k=-1$, $\Sigma$
is a hyperbolic hypersurface. By an appropriate identification, in this case,
one may get an arbitrary higher genus hypersurface. In the asymptotically
anti-de Sitter spaces, black holes with these three horizon hypersurfaces
exist. Now we discuss these so-called topological black holes in the
dilaton gravity described by (\ref{4e1}). Because of the dilaton
potential, we will see that the topological dilaton black holes will
not approach asymptotically the anti-de Sitter space. Let us consider the
case $k=0$ first.

\subsection{$k=0$ solutions}

In the case of $k=0$, we have the black hole solutions \cite{Cai1,Cai2}:
\begin{eqnarray}
\label{4e3}
&& A(r) =-\frac{8\pi Gm}{VN\beta^2}r^{1-2N} +
  \frac{\Lambda e^{2b\phi_0}}{N(4N-1)}r^{2N}
+ \frac{16\pi^2 Q^2 e^{2a\phi_0}}{NV^2\beta^4}r^{-2N}, \nonumber\\
&& R(r) =\beta r^{N}, \nonumber \\
&& \phi= \phi_0 -\sqrt{N(1-N)}\ln\ r, \nonumber\\
&& F_{tr}=\frac{4\pi Q}{VR^2}e^{2a\phi}, \nonumber\\
&& a=b =\sqrt{N(1-N)}/N,
\end{eqnarray}
where $\phi_0$ is an integration constant, $Q$ is the charge of the hole
and $m$ is  the quasilocal mass identified at spatial infinity. $N$ and $\beta$
are two parameters and $V$ is the area of the hypersurface
$\Sigma$. In order for the solution (\ref{4e3}) to have a black hole
structure, it must be satisfied  that $\Lambda >0$ and $ 1/4<N<1$ \cite{Cai1}.

For the solution (\ref{4e3}), the results in the appendix tell us that
the suitable surface counterterm is
\begin{equation}
\label{4e4}
S_{\rm ct}=-\frac{1}{8\pi G}\int d^3x\sqrt{-h}\frac{c_0}{l_{\rm eff}},
\ \ \  c_0= 2 \sqrt{\frac{3N}{4N-1}}, \ \ \ \frac{1}{l_{\rm eff}}=
  e^{b\phi}\sqrt{\frac{\Lambda}{3}}.
\label{c4}
\end{equation}
Using this surface counterterm, we obtain the following quasilocal
stress-energy tensor:
\begin{eqnarray}
&&  T_{tt}= \frac{ m}{V\beta^2}c_1r^{-N} + \cdots, \nonumber\\
&&   T_{xx}= \frac{m }{2Vc_1}\frac{2N-1}{N} r^{-N} +\cdots,
       \nonumber \\
&& T_{yy}=T_{xx},
\end{eqnarray}
where $c_1^2=\Lambda e^{2b\phi_0}/N(4N-1)$. According to the mass
formula (\ref{1e6}), the mass of the black hole is
\begin{equation}
M=\int_{r\rightarrow \infty} d^2x\sqrt{\sigma} R^2 A^{-1/2}T_{tt}=m,
\end{equation}
where $\sigma$ is the determinant of the metric of the hypersurface $\Sigma$.
The mass is the same as the quasilocal mass at the spatial infinity. The
surface metric, in which the boundary quantum field lives, is
\begin{equation}
\gamma_{ab}=\lim_{r\rightarrow \infty}\frac{1}{R^2}h_{ab}
     =- \beta^{-2}c_1^2 dt^2 +dx^2 +dy^2.
\end{equation}
Using (\ref{2e21}), we then obtain the boundary stress-energy tensor of
the corresponding boundary quantum field,
\begin{equation}
\tau_{ab}=\frac{M}{Vc_1}\frac{2N-1}{2N}
    {\rm diag}\left [\frac{2N}{2N-1}\frac{c_1^2}{\beta^2},
    1,1\right].
\end{equation}
After a straightforward calculation using the counterterm (\ref{4e4}), we
are able to get a finite Gibbs free energy of the black hole:
\begin{equation}
\label{4e9}
{\cal G}=-\frac{(2N-1)V\beta^2}{16\pi G N}\left [\frac{\Lambda e^{2b\phi_0}}
   {4N-1} r_+^{4N-1} +\frac{16\pi^2 Q^2 e^{2a\phi_0}}{V^2 \beta^4}
   \frac{1}{r_+}\right],
\end{equation}
where $r_+$ is the horizon of the black hole, which satisfies
the equation $ A(r_+)=0$. The Gibbs free energy (\ref{4e9}) is consistent
with the definition
\begin{equation}
\label{definition}
{\cal G}= M-TS-\mu Q,
\end{equation}
where $T$ and $S$ are the Hawking temperature and entropy of black holes, and
$\mu$ is the chemical potential corresponding to the charge. For the solution
(\ref{4e3}), we have
\begin{eqnarray}
\label{entropy}
&& T=\frac{1}{4\pi r_+}\left [\frac{\Lambda e^{2b\phi_0}}{N}r_+^{2N}
     -\frac{16\pi^2 Q^2e^{2a\phi_0}}{NV^2\beta^4}r_+^{-2N}\right],
   \nonumber\\
&& S=\frac{V\beta^2}{4G}r_+^{2N},\ \ \ \mu=\frac{4\pi Q e^{2a\phi_0}}
     {GV\beta^2 r_+}.
\end{eqnarray}
It is easy to verify that (\ref{definition}) reproduces the result
(\ref{4e9}). Because of $ 1/4 <N<1$, we see from (\ref{4e9}) that
the free energy is always negative for $1/2 <N<1$, but changes
its sign at $N=1/2$, becoming positive for $1/4 <N<1/2$. In the latter case,
the black hole is thermodynamically unstable. Note that in the Einstein
\cite{Bir} and Einstein-Maxwell \cite{Cai3,Cham2} gravities with a negative
cosmological constant, the black holes with $k=0$ are always
thermodynamically stable. Therefore the change of asymptotic behavior of
black hole solutions may change the thermodynamic stability. It is
worth pointing out here that the sign change of the Gibbs free energy
does not mean the occurrence of the Hawking-Page phase transition
\cite{Page,Witten2} as in the Einstein gravity, because once given a black
hole, the value $N$ is fixed and then the sign of the free energy is
fixed as well and will not change due to the change of the size of
black holes.

\subsection{ $k=-1$ solutions}

In this case, we consider the following black hole solution \cite{Cai1}:
\begin{eqnarray}
\label{4e10}
&& A(r)=-\frac{8\pi Gm}{VN \beta^2 }r^{1-2N} +\frac{\Lambda e^{2b\phi_0}}
    {1-N}r^{2-2N} +\frac{16\pi^2 Q^2 e^{2a\phi_0}}{NV^2 \beta^4}r^{-2N},
     \nonumber \\
&& \phi =\phi_0 -\sqrt{N(1-N)}\ln\ r, \nonumber\\
&& \Lambda = \frac{1-N}{1-2N}\frac{e^{-2b\phi_0}}{\beta^2}, \nonumber \\
&& b=1/a= N/\sqrt{N(1-N)},
\end{eqnarray}
where we use the same notations as in the previous subsection and in
particular $R$ and $F_{tr}$ are the same. To have a black hole structure,
we must have $0<N<1/2$.

For the hyperbolic black hole (\ref{4e10}), we find that the appropriate
surface counterterm is
\begin{equation}
\label{4e11}
S_{\rm ct} =-\frac{1}{8\pi G}\int d^3x\sqrt{-h}\frac{c_0}{l_{\rm eff}},
 \ \ \  c_0=2N\sqrt{\frac{3}{1-N}},\ \ \ \frac{1}{l_{\rm eff}}=
 e^{b\phi}\sqrt{\frac{\Lambda}{3}}.
\end{equation}
The quasilocal stress-energy tensor then is
\begin{equation}
T_{tt}=\frac{m e^{b\phi_0}}{V\beta^2}\sqrt{\frac{\Lambda}{1-N}}
  r^{1-3N} +\cdots.
\end{equation}
Unfortunately, the black hole solution (\ref{4e10}) has no well-defined
surface metric $\gamma_{ab}$ in this case and the other components of the
quasilocal stress-energy tensor are not well defined, either. Nonetheless,
as a consistency check of the surface counterterm (\ref{4e11}), we may
compute the mass of the black hole using (\ref{1e6}). Once again, in this
way, the mass of the black hole is found to coincide with the quasilocal
mass $m$ at the spatial infinity:
\begin{equation}
M=\int_{r\rightarrow \infty}d^2x \sqrt{\sigma}R^2 A^{-1/2}T_{tt}=m.
\end{equation}
Note that due to the different asymptotic behavior, the so-called
``negative mass'' black holes \cite{Bir,Cai3,Em} do not appear in this
dilaton gravity.
The Euclidean action of the hole is
\begin{eqnarray}
I= &-&\frac{1}{16\pi G}\int d^4x\sqrt{g}[ R-2(\partial\phi)^2
    +2\Lambda e^{2b\phi} -e^{-2a\phi}F_{\mu\nu}F^{\mu\nu}]
  \nonumber \\
 & +& \frac{1}{8\pi G}\int d^3x\sqrt{h}K
+\frac{1}{8\pi G}\int d^3x\sqrt{h}\frac{c_0}{l_{\rm eff}}.
\end{eqnarray}
Substituting the solution and the ``effective cosmological
constant'' into the above, we find
\begin{equation}
\label{4e15}
I\equiv {\cal G}/T =\frac{V}{16\pi G T}\left [
 -r_+ +\frac{1-2N}{N}\frac{16\pi^2 Q^2 e^{2a\phi_0}}
    {V^2 \beta^2 r_+}\right ],
\end{equation}
where $T$ is the Hawking temperature of the solution,
\begin{equation}
T=\frac{1}{4\pi r_+}\left [\frac{\Lambda}{1-N}e^{2b\phi _0}r_+^{2-2N}
   -\frac{16\pi^2Q^2}{NV^2\beta^2}e^{2a\phi_0}r_+^{-2N}\right],
\end{equation}
and $r_+$ is the horizon radius obeying the equation
\begin{equation}
 \frac{1}{1-2N}r_+^2 -\frac{8\pi Gm}{VN }r_+ +
    \frac{16\pi^2 Q^2 e^{2a\phi_0}}{NV^2 \beta^2}=0.
\end{equation}
{}From (\ref{4e15}) we see that the first term is negative, while the second
is positive. Therefore the Gibbs free energy may be negative for large black
holes, while positive for small black holes, which is reminiscent of the
Schwarzschild-anti-de Sitter black holes \cite{Page,Witten2}, where
the free energy is also negative for large black holes
and positive for small black holes. Therefore the Hawking-Page phase
transition  takes place for the hyperbolic black holes (\ref{4e10}), which
occurs at ${\cal G}=0$, that is, at $r=r_+$ with
\begin{equation}
\label{crit}
r_+^2 =\frac{1-2N}{N}\frac{16\pi^2 Q^2e^{2a\phi_0}}{V^2\beta^2}.
\end{equation}
{}From the free energy (\ref{4e15}) and the critical point (\ref{crit}),
one may see that the charge plays a central role in the Hawking-Page
phase transition: if $Q=0$, the phase transition will disappear. It would
be interesting to note that the Hawking-Page phase transition does not
appear in the hyperbolic black holes of the
 Einstein and Einstein-Maxwell AdS gravities \cite{Bir,Cai3,Em}. In addition,
as a consistency check, one may reproduce  the Gibbs free energy (\ref{4e15})
by the definition (\ref{definition}) with the same expressions in
(\ref{entropy}) of black hole entropy $S$ and the
chemical potential $\mu$.

\subsection{ $k=1$ solutions}

Of course, the action (\ref{4e1}) has also spherically symmetric
black hole solutions. Three sets of black hole solutions have
been given in \cite{CHM}. We consider here the first set of the solutions
found there:
\begin{eqnarray}
\label{4e17}
&& A(r)= r^{\frac{2a^2}{1+a^2}}\left (\frac{1+a^2}{(1-a^2)\beta^2}
    -\frac{2(1+a^2)Gm}{\beta^2 r} +\frac{Q^2 (1+a^2)e^{2a\phi_0}}
       {\beta^4 r^2}\right), \nonumber \\
&& R^2(r)=\beta^2 r^{2N}, \ \ \ N=1/(1+a^2), \nonumber\\
&& \phi= \phi_0 -\frac{a}{1+a^2}\ln\ r, \nonumber\\
&& F_{tr}=\frac{Q e^{2a\phi}}{R^2}, \nonumber\\
&&  b=1/a,\ \ \ \   \Lambda = \frac{a^2}{(1-a^2)\beta^2}e^{-2\phi_0/a}.
\end{eqnarray}
For this solution, $a^2<1$ must be satisfied in order for the solution
to describe a black hole. In addition, one may notice that when
$a^2 \rightarrow 0$, the solution has a well-defined asymptotically flat
limit: Reissner-Nordstr\"om (RN) black hole solution.

As described in the appendix, the appropriate surface counterterm is
\begin{equation}
\label{4ect}
S= -\frac{1}{8\pi G}\int d^3x \sqrt{-h}\frac{c_0}{l_{\rm eff}},\ \ \
  c_0=2\sqrt{\frac{3}{a^2(1+a^2)}},\ \ \ \frac{1}{l_{\rm eff}}=e^{b\phi}
  \sqrt{\frac{\Lambda}{3}}.
\end{equation}
Using this surface term, we have
\begin{equation}
T_{tt}=\frac{m}{4\pi \beta^2 }\sqrt{\frac{1+a^2}{(1-a^2)\beta^2}}
   r^{\frac{a^2-2}{a^2+1}} +\cdots.
\end{equation}
A well-defined surface metric $\gamma_{ab}$ requires $a^2=1$, but which is
excluded by the existence of black hole solutions from (\ref{4e17}).
Just as the case of $k=-1$, we cannot obtain a well-defined stress-energy
tensor of the corresponding quantum field here. Instead we can calculate
the mass of the black hole as before. The result is again in agreement
with the quasilocal mass $m$ identified at the spatial infinity,
\begin{equation}
M=\int_{r\rightarrow \infty} d\theta d\varphi \sin^2\theta R^2 A^{-1/2}T_{tt}
 =m.
\end{equation}
The Euclidean action of the black hole in the grand canonical ensemble
is found to be
\begin{equation}
\label{4e22}
I \equiv {\cal G}/T =\frac{1}{4GT}\left [ r_+ -(1-a^2) \frac{Q^2 e^{2a\phi_0}}
    {\beta^2 r_+}\right].
\end{equation}
Here $T$ is the Hawking temperature of the black hole
\begin{equation}
T=\frac{1}{4\pi}r_+^{\frac{a^2-1}{a^2+1}}\left [\frac{1+a^2}{(1-a^2)\beta^2}
    -\frac{Q^2(1+a^2)e^{2a\phi_0}}{\beta^4 r_+^2}\right],
\end{equation}
and $r_+$ is the horizon radius satisfying the equation
\begin{equation}
\frac{1}{(1-a^2)}
    -\frac{2Gm}{ r_+} +\frac{Q^2 e^{2a\phi_0}}
       {\beta^2 r_+^2}=0.
\end{equation}
It can be seen clearly from (\ref{4e22}) that the free energy is negative
for small black holes, while it becomes positive for large black holes, which
changes its sign at
\begin{equation}
r_+^2 =(1-a^2)\frac{Q^2e^{2a\phi_0}}{\beta^2}.
\end{equation}
Thus the small black holes are thermodynamically stable and large black holes
will become thermodynamically unstable. This property is the same as that of
the RN black holes in asymptotically flat spaces. As is well known, the heat
capacity is positive for near-extremal RN black holes (small $r_+$) and
becomes negative beyond a certain critical point ($r_+$ gets
larger) from extremal RN black holes. This thermodynamic behavior is
completely opposite to that of the black holes in the anti-de Sitter
spaces \cite{Page}. Therefore although an ``effective negative cosmological
constant'' occurs here, the thermodynamic properties are similar to those
of RN black holes in asymptotically flat spaces. Furthermore, we can check
that the Gibbs free energy (\ref{4e22}) is reproduced by the definition
with the black hole entropy $S$ and the chemical potential $\mu$:
\begin{equation}
S=\frac{\pi \beta^2}{G}r_+^{\frac{2}{1+a^2}},\ \ \  \mu=\frac{Qe^{2a\phi_0}}
  {G\beta^2 r_+}.
\end{equation}
In addition, it
is worth noting that because the solution ({\ref{4e17}) has a well-defined
asymptotically flat limit as $a\rightarrow 0$, the surface counterterm
(\ref{4ect}) has also a well-defined asymptotically flat limit. In this
limit, we again reproduce the thermodynamics of RN and Schwarzschild black
holes.


\sect{Conclusions}

In the dilaton gravities with a dilaton potential, in general,
 the black hole solutions do  not approach asymptotically
anti-de Sitter spaces due to the dilaton field, but we have found that for
such black holes, it is also possible to extract a well-defined surface
stress-energy tensor and to get finite Euclidean action by adding
appropriate surface counterterms to the bulk action, in which the
dilaton potential plays a similar role as the cosmological constant does
in the Einstein(-Maxwell) gravity. In this paper using this prescription
we studied some examples including domain-wall black holes in gauged
supergravities, three-dimensional dilaton black holes and topological
dilaton black holes in four dimensions. In these examples, this prescription
works well.

Using the surface counterterm method, we have also obtained boundary
stress-energy tensors and Euclidean actions of domain-wall black holes.
These results have been checked to be consistent with those coming from
direct calculations in the original D$p$-brane configurations of type II
supergravity. For a kind of charged domain-wall black holes in the
domain-wall gauged supergravities, which result from singular limit of the
sphere reduction of eleven-dimensional supergravity and ten-dimensional type
IIB supergravity, we  have found that the Gibbs free energy and the pressure
of the thermal excitations on the domain-wall always vanish. This is the
same as the situation for the D5-brane case.

We have also studied thermodynamics of these black configurations by
calculating Euclidean action within this surface counterterm method.
Some  new features have been found in the topological dilaton black holes,
which are not present in the Einstein(-Maxwell) gravities with a negative
cosmological constant. For example, $k=0$ dilaton black holes may
be thermodynamically unstable; in the hyperbolic charged dilaton black
holes ($k=-1$), the Hawking-Page phase transition may take place; in the
case of $k=1$ we have a well-defined asymptotically flat limit of the
surface counterterm. Using it, we can reproduce the thermodynamics of
Schwarzschild and RN black holes.

\section* {Acknowledgments}

This work was supported in part by the Japan Society for the Promotion of
Science and by grant-in-aid from the Ministry of Education, Science,
Sports and Culture No. 99020.

\appendix

\sect{Formula for surface counterterm}

Here we present our formula for the surface counterterm, which is applicable
to all our cases.

Let us write our metric as
\begin{equation}
ds^2 = - A(r) dt^2 + B(r) dr^2 + C(r) dx_p^2,
\end{equation}
where the last term represents a $p$-dimensional Ricci-flat space,
and the ``effective cosmological constant" $l_{\rm eff}$ is defined as
\begin{equation}
\frac{1}{l_{\rm eff}} = \sqrt{\frac{V(\Phi)}{p(p+1)}},
\end{equation}
in the Einstein frame. By analogy with the surface counterterms in
Eq.~(\ref{1e2}), we introduce the counterterm
\begin{equation}
-2 \int d^{p+1}x \sqrt{-h} \frac{c_0}{l_{\rm eff}},
\label{for1}
\end{equation}
where only the relative normalization with the Einstein term is written
explicitly. In fact, we find that appropriate choice of the coefficient
removes divergences from physical quantities. The asymptotic behaviors of
the metrics $B,C$ and $V$ govern the coefficient $c_0$, whose formula is
derived below.

Let the asymptotic behaviors of the fields be
\begin{eqnarray}
A(r) &=& A_0 r^\alpha + \ldots, \nonumber\\
B(r) &=& B_0 r^\beta + \ldots, \nonumber\\
C(r) &=& C_0 r^\gamma + \ldots, \nonumber\\
V(\Phi) &=& V_0 r^\delta + \ldots.
\end{eqnarray}
It can be easily checked that in order to satisfy the Einstein equations,
one must have
\begin{equation}
\beta + \delta = -2.
\label{cond}
\end{equation}
We then compute the boundary stress-energy tensor to obtain
\begin{eqnarray}
T_{tt} &=& - \frac{p A}{2\sqrt{B}}\frac{C'}{C} + \frac{c_0}{l_{\rm eff}} A,
\nonumber\\
&\simeq& - \frac{p A_0 \gamma}{2 \sqrt{B_0}} r^{\alpha -\beta/2 -1}
+ \frac{c A_0 \sqrt{V_0}}{\sqrt{p(p+1)}} r^{\alpha + \delta/2} +\ldots ,
\label{beh}
\end{eqnarray}
where prime indicates differentiation with respect to $r$. This leading
behavior governs the finiteness of the physical quantities. Note that
thanks to the relation (\ref{cond}), the two terms in Eq.~(\ref{beh}) match
with each other and allows to cancel the divergences.

Imposing the condition that the leading terms be absent in Eq.~(\ref{beh}),
we obtain the formula
\begin{equation}
c_0 = \frac{\gamma p\sqrt{p(p+1)}}{2\sqrt{B_0 V_0}}.
\label{for2}
\end{equation}
In fact, for the asymptotically AdS space
\begin{equation}
\gamma=2, \;\;
B_0 = l^2,\;\;
V_0 = \frac{p(p+1)}{l^2},
\end{equation}
which reproduces the first term in Eq.~(\ref{1e2}).

For the neutral domain-wall black holes, we find
\begin{equation}
\gamma = \frac{2(p-9)}{p(p-5)}, \;\;
B_0 V_0 = \frac{2(9-p)(7-p)}{(5-p)^2},
\label{a1}
\end{equation}
giving (\ref{b1}). For the charged domain-wall, we have
\begin{equation}
\gamma =1, \;\;
B_0 V_0 = \frac{p^2}{4},
\label{a2}
\end{equation}
leading to (\ref{b2}). For the dilaton black holes,
\begin{equation}
\gamma =N, \;\;
p=1, \;\;
B_0 V_0 = \frac{N(3N-2)}{4},
\label{a3}
\end{equation}
yielding (\ref{b3}). For topological black holes, we find
\begin{equation}
\gamma =2N, \;\;
p=2,
\label{a4}
\end{equation}
and
\begin{eqnarray}
&& B_0 V_0 = 2N(4N-1),\;\; {\rm for}\; k=0, \nonumber\\
&& B_0 V_0 = 2(1-N),\;\; {\rm for}\; k=-1, \nonumber\\
&& B_0 V_0 = \frac{2a^2}{1+a^2},\;\; {\rm for}\; k=1,
\label{a5}
\end{eqnarray}
giving (\ref{c4}), (\ref{4e11}) and (\ref{4ect}), respectively.


\end{document}